\documentclass[preprint,superscriptaddress,nofootinbib]{revtex4}
  \usepackage{graphicx}
  \begin{document}
  \title{${\Upsilon}(1S)$ ${\to}$ $B_{c}{\pi}$, $B_{c}K$
         decays with perturbative QCD approach}
  \author{Junfeng Sun}
  \affiliation{Institute of Particle and Nuclear Physics,
              Henan Normal University, Xinxiang 453007, China}
  \author{Qingxia Li}
  \affiliation{Institute of Particle and Nuclear Physics,
              Henan Normal University, Xinxiang 453007, China}
  \author{Yueling Yang}
  \affiliation{Institute of Particle and Nuclear Physics,
              Henan Normal University, Xinxiang 453007, China}
  \author{Haiyan Li}
  \affiliation{Institute of Particle and Nuclear Physics,
              Henan Normal University, Xinxiang 453007, China}
  \author{Qin Chang}
  \affiliation{Institute of Particle and Nuclear Physics,
              Henan Normal University, Xinxiang 453007, China}
  \author{Zhiqing Zhang}
  \affiliation{Department of Physics, Henan University of Technology,
               Zhengzhou 450001, China}

  \begin{abstract}
  With the potential prospects of the ${\Upsilon}(1S)$
  at high-luminosity dedicated heavy-flavor factories,
  the bottom-changing ${\Upsilon}(1S)$ ${\to}$ $B_{c}{\pi}$,
  $B_{c}K$ weak decays are studied with the pQCD approach.
  It is found that branching ratio for the color-favored and
  CKM-favored ${\Upsilon}(1S)$ ${\to}$ $B_{c}{\pi}$ decay
  can reach up to ${\cal O}(10^{-11})$. So
  the ${\Upsilon}(1S)$ ${\to}$ $B_{c}{\pi}$ decay might be
  measured promisingly by the future experiments.
  \end{abstract}
  \maketitle

  \section{Introduction}
  \label{sec01}
  The ${\Upsilon}(1S)$ particle is the ground vector
  bottomonia (bound states of $b\bar{b}$) with well
  established quantum number of $I^{G}J^{PC}$ $=$
  $0^{-}1^{--}$ \cite{pdg}.
  The mass of the ${\Upsilon}(1S)$ particle,
  $m_{{\Upsilon}(1S)}$ $=$ $9460.30{\pm}0.26$ MeV \cite{pdg},
  is less than the kinematic $B\bar{B}$ threshold.
  The ${\Upsilon}(1S)$ particle,
  in a close analogy with $J/{\psi}$,
  decay primarily through the annihilation of the
  $b\bar{b}$ pairs into three gluons, followed by
  evolution of gluons into hadrons, glueballs,
  hybrid, multiquark and other exotic states.
  The hadronic ${\Upsilon}(1S)$ decay offers an ideal place
  to reap the properties of the invisible gluons and
  of the quark-gluon coupling \cite{ann1983}.
  It is well known that strong decay of the
  ${\Upsilon}(1S)$ particle is suppressed by the
  phenomenological Okubo-Zweig-Iizuka (OZI)
  rules \cite{ozi-o,ozi-z,ozi-i},
  so electromagnetic and radiative transitions
  become competitive.
  Besides, the ${\Upsilon}(1S)$ particle can also decay
  via the weak interactions within the standard model,
  although the branching ratio is very small,
  about $2/{\tau}_{B}{\Gamma}_{{\Upsilon}(1S)}$
  ${\sim}$ ${\cal O}(10^{-8})$ \cite{pdg}, where
  ${\tau}_{B}$ and ${\Gamma}_{{\Upsilon}(1S)}$ are
  the lifetime of the $B_{u,d,s}$ meson and the decay
  width of the ${\Upsilon}(1S)$ particle, respectively.
  In this paper, we will estimate the branching ratios
  for the bottom-changing nonleptonic ${\Upsilon}(1S)$
  ${\to}$ $B_{c}{\pi}$, $B_{c}K$ weak decays with
  perturbative QCD (pQCD) approach
  \cite{pqcd1,pqcd2,pqcd3}.
  The motivation is listed as follows.

  From the experimental point of view,
  (1)
  over $10^{8}$ ${\Upsilon}(1S)$ samples have been
  accumulated at Belle \cite{1212.5342}.
  Many more upsilons could be collected with great
  precision at the forthcoming SuperKEKB and the
  running upgraded LHC.
  The abundant ${\Upsilon}(1S)$ samples provide a golden
  opportunity to search for the ${\Upsilon}(1S)$
  weak decays which in some cases might be detectable.
  Theoretical studies on the ${\Upsilon}(1S)$
  weak decays are just necessary to offer a ready
  reference.
  (2)
  For the two-body ${\Upsilon}(1S)$ ${\to}$ $B_{c}{\pi}$,
  $B_{c}K$ decays, final states with opposite charges
  have definite energies and momenta in the rest
  frame of the ${\Upsilon}(1S)$ particle.
  Besides, identification of a single explicitly
  flavored $B_{c}$ meson is free from
  inefficiently double tagging above the
  $B\bar{B}$ threshold \cite{zpc62.271},
  and can provide a conclusive evidence of the
  ${\Upsilon}(1S)$ weak decay.
  Of course, small branching ratios make the observation
  of the ${\Upsilon}(1S)$ weak decays extremely challenging,
  and evidences of an abnormally large production rate
  of single $B_{c}$ mesons in the ${\Upsilon}(1S)$ decay
  might be a hint of new physics \cite{zpc62.271}.

  From the theoretical point of view,
  the bottom-changing upsilon weak decays permit one
  to cross check parameters obtained from $B$ meson decay.
  The color-favored
  ${\Upsilon}(1S)$ ${\to}$ $B_{c}{\pi}$, $B_{c}K$
  decays have been estimated with the naive factorization
  (NF) approximation in previous works
  \cite{zpc62.271,ijma14,adv2013}.
  An obvious deficiency of NF approximation is the
  absence of strong phases and the renormalization
  scale from hadronic matrix elements (HME).
  Recently, several attractive methods have been
  developed to evaluate HME,
  such as pQCD \cite{pqcd1,pqcd2,pqcd3},
  the QCD factorization (QCDF) \cite{qcdf1,qcdf2,qcdf3}
  and soft and collinear effective theory
  \cite{scet1,scet2,scet3,scet4},
  which could give reasonable explanation for
  many measurements on $B_{u,d}$ hadronic decays.
  The ${\Upsilon}(1S)$ ${\to}$ $B_{c}{\pi}$, $B_{c}K$
  decays are calculated at the next-to-leading (NLO)
  order with the QCDF approach \cite{691261}.
  In this paper, the ${\Upsilon}(1S)$ ${\to}$
  $B_{c}{\pi}$, $B_{c}K$ weak decays will be
  evaluated with the pQCD approach to check the
  consistency of prediction on branching ratios
  among different models.

  This paper is organized as follows.
  In section \ref{sec02}, we present the theoretical framework
  and the amplitudes for the ${\Upsilon}(1S)$ ${\to}$ $B_{c}{\pi}$,
  $B_{c}K$ decays with the pQCD approach.
  Section \ref{sec03} is devoted to numerical results and discussion.
  The last section is our summary.

  \section{theoretical framework}
  \label{sec02}
  \subsection{The effective Hamiltonian}
  \label{sec0201}
  The effective Hamiltonian responsible for the
  ${\Upsilon}(1S)$ ${\to}$ $B_{c}{\pi}$, $B_{c}K$
  decays is \cite{9512380}
   \begin{equation}
  {\cal H}_{\rm eff}\ =\ \frac{G_{F}}{\sqrt{2}}\,
   \sum\limits_{q=d,s}\, V_{cb} V_{uq}^{\ast}\,
   \Big\{ C_{1}({\mu})\,Q_{1}({\mu})
         +C_{2}({\mu})\,Q_{2}({\mu}) \Big\}
   + {\rm H.c.}
   \label{hamilton},
   \end{equation}
  where $G_{F}$ $=$ $1.166{\times}10^{-5}\,{\rm GeV}^{-2}$ \cite{pdg}
  is the Fermi coupling constant;
  the Cabibbo-Kabayashi-Maskawa (CKM) factors
  are expanded as a power series in
  the Wolfenstein parameter ${\lambda}$ ${\sim}$ $0.2$ \cite{pdg},
  \begin{eqnarray}
  V_{cb}V_{ud}^{\ast} &=&
               A{\lambda}^{2}
  - \frac{1}{2}A{\lambda}^{4}
  - \frac{1}{8}A{\lambda}^{6}
  +{\cal O}({\lambda}^{8})
  \label{eq:ckm01}, \\
  V_{cb}V_{us}^{\ast} &=& A{\lambda}^{3}
  +{\cal O}({\lambda}^{8})
  \label{eq:ckm02}.
  \end{eqnarray}
  The Wilson coefficients $C_{1,2}(\mu)$ summarize the
  physical contributions above scales of ${\mu}$,
  and have properly been calculated to the NLO order
  with the renormalization group improved perturbation
  theory.
  The local operators are defined as follows.
    \begin{eqnarray}
    Q_{1} &=&
  [ \bar{c}_{\alpha}{\gamma}_{\mu}(1-{\gamma}_{5})b_{\alpha} ]
  [ \bar{q}_{\beta} {\gamma}^{\mu}(1-{\gamma}_{5})u_{\beta} ]
    \label{q1}, \\
    Q_{2} &=&
  [ \bar{c}_{\alpha}{\gamma}_{\mu}(1-{\gamma}_{5})b_{\beta} ]
  [ \bar{q}_{\beta}{\gamma}^{\mu}(1-{\gamma}_{5})u_{\alpha} ]
    \label{q2},
    \end{eqnarray}
  where ${\alpha}$ and ${\beta}$ are color indices and the
  sum over repeated indices is understood.

  From the effective Hamiltonian Eq.(\ref{hamilton}),
  it can be easily seen that only tree operators
  with coupling strength proportional to the CKM
  element $V_{cb}$
  contribute to the ${\Upsilon}(1S)$ ${\to}$
  $B_{c}{\pi}$, $B_{c}K$ decays, and there is no
  pollution from penguin and annihilation
  contributions.

  \subsection{Hadronic matrix elements}
  \label{sec0202}
  To obtain the decay amplitudes, the remaining works are
  how to calculate accurately hadronic matrix elements
  of local operators.
  Using the Lepage-Brodsky approach for exclusive
  processes \cite{prd22}, HME
  could be expressed as the convolution of hard scattering
  subamplitudes containing perturbative contributions
  with the universal wave functions reflecting the
  nonperturbative contributions.
  Sometimes the high-order corrections to HME
  produce collinear and/or soft logarithms
  based on collinear factorization approximation,
  for example, the spectator scattering amplitudes
  within the QCDF framework \cite{qcdf3}.
  The pQCD approach advocates
  that \cite{pqcd1,pqcd2,pqcd3} this inconsistent
  treatment on HME could be smeared by retaining
  the transverse momentum of quarks and
  introducing the Sudakov factor.
  The decay amplitudes could be factorized into three
  parts: the ``harder'' effects incorporated into
  the Wilson coefficients $C_{i}$, the process-dependent
  heavy quark decay subamplitudes $H$, and the
  universal wave functions ${\Phi}$; and are
  written as
  \begin{equation}
  {\int} dx\, db\,
  C_{i}(t)H(t,x,b){\Phi}(x,b)e^{-S}
  \label{hadronic},
  \end{equation}
  where $t$ is a typical scale, $x$ is the
  longitudinal momentum fraction of the valence quark,
  $b$ is the conjugate variable of the transverse
  momentum, and $e^{-S}$ is the Sudakov factor.

  \subsection{Kinematic variables}
  \label{sec0203}
  The light cone kinematic variables in the ${\Upsilon}(1S)$
  rest frame are defined as follows.
  \begin{equation}
  p_{{\Upsilon}}\, =\, p_{1}\, =\, \frac{m_{1}}{\sqrt{2}}(1,1,0)
  \label{kine-p1},
  \end{equation}
  \begin{equation}
  p_{B_{c}}\, =\, p_{2}\, =\, (p_{2}^{+},p_{2}^{-},0)
  \label{kine-p2},
  \end{equation}
  \begin{equation}
  p_{\pi(K)}\, =\, p_{3}\, =\, (p_{3}^{-},p_{3}^{+},0)
  \label{kine-p3},
  \end{equation}
  \begin{equation}
  k_{i}\, =\, x_{i}\,p_{i}+(0,0,\vec{k}_{i{\perp}})
  \label{kine-ki},
  \end{equation}
  \begin{equation}
  {\epsilon}_{\Upsilon}^{\parallel}\, =\, \frac{1}{ \sqrt{2} }(1,-1,0)
  \label{kine-1el},
  \end{equation}
  \begin{equation}
  n_{+}=(1,0,0), \quad n_{-}=(0,1,0)
  \label{kine-null},
  \end{equation}
  where $x_{i}$ and $\vec{k}_{i{\perp}}$
  are the longitudinal momentum fraction and
  transverse momentum of the light valence quark,
  respectively; ${\epsilon}_{\Upsilon}^{\parallel}$
  is the longitudinal polarization vector
  of the ${\Upsilon}(1S)$ particle;
  $n_{+}$ and $n_{-}$ are the positive and negative
  null vectors, respectively.
  The notation of momentum is displayed
  in Fig.\ref{fig1}(a).

  The relations among these kinematic variables are
  \begin{equation}
  p_{i}^{\pm}\, =\, \frac{E_{i}\,{\pm}\,p}{\sqrt{2}}
  \label{kine-pipm},
  \end{equation}
  \begin{equation}
  p\, =\, \frac{\sqrt{ {\lambda}(m_{1}^{2},m_{2}^{2},m_{3}^{2}) }}{2\,m_{1}}
  \label{kine-pcm},
  \end{equation}
  \begin{equation}
  s\, =\, 2\,p_{2}{\cdot}p_{3}\, =\ m_{1}^{2}-m_{2}^{2}-m_{3}^{2}
  \label{kine-s},
  \end{equation}
  \begin{equation}
  t\, =\, 2\,p_{1}{\cdot}p_{2}\, =\ m_{1}^{2}+m_{2}^{2}-m_{3}^{2}
  \label{kine-t},
  \end{equation}
  \begin{equation}
  u\, =\, 2\,p_{1}{\cdot}p_{3}\, =\ m_{1}^{2}-m_{2}^{2}+m_{3}^{2}
  \label{kine-u},
  \end{equation}
  \begin{equation}
  t+u-s\, =\, m_{1}^{2}+m_{2}^{2}+m_{3}^{2}
  \label{kine-stu01},
  \end{equation}
  \begin{equation}
  s\,t+s\,u-u\,t\, =\, {\lambda}(m_{1}^{2},m_{2}^{2},m_{3}^{2})
  \label{kine-stu02},
  \end{equation}
  \begin{equation}
  {\lambda}(a,b,c)\, =\, a^{2}+b^{2}+c^{2}-2\,a\,b-2\,a\,c-2\,b\,c
  \label{kine-lambda},
  \end{equation}
  where $p$ is the common momentum of final states;
  $m_{1}$, $m_{2}$ and $m_{3}$ denote the masses
  of the ${\Upsilon}(1S)$, $B_{c}$ and ${\pi}(K)$
  mesons, respectively.

  \subsection{Wave functions}
  \label{sec0204}
  With the notation in
  \cite{prd65.014007,npb529.323,jhep0605.004},
  the explicit definitions of matrix elements of
  diquark operators sandwiched between
  vacuum and the longitudinally polarized
  ${\Upsilon}(1S)$,
  the double-heavy pseudoscalar $B_{c}$,
  the light pseudoscalar $P$ ($=$ ${\pi}$, $K$) are
  \begin{equation}
 {\langle}0{\vert}b_{i}(z)\bar{b}_{j}(0){\vert}
 {\Upsilon}(p_{1},{\epsilon}_{\parallel}){\rangle}\,
 =\, \frac{1}{4}f_{\Upsilon}
 {\int}dx_{1}\,e^{-ix_{1}p_{1}{\cdot}z}
  \Big\{ \!\!\not{\epsilon}_{\parallel} \Big[
   m_{1}\,{\phi}_{\Upsilon}^{v}(x_{1})
  -\!\!\not{p}_{1}\, {\phi}_{\Upsilon}^{t}(x_{1})
  \Big] \Big\}_{ji}
  \label{wave-bbl},
  \end{equation}
  \begin{equation}
 {\langle}B_{c}^{+}(p_{2}){\vert}\bar{c}_{i}(z)b_{j}(0){\vert}0{\rangle}\,
 =\, \frac{i}{4}f_{B_{c}} {\int}dx_{2}\,e^{ix_{2}p_{2}{\cdot}z}\,
  \Big\{ {\gamma}_{5}\Big[ \!\!\not{p}_{2}+m_{2}\Big]
 {\phi}_{B_{c}}(x_{2}) \Big\}_{ji}
  \label{wave-bcp},
  \end{equation}
  \begin{eqnarray}
  & &
 {\langle}P(p_{3}){\vert}u_{i}(z)\bar{q}_{j}(0){\vert}0{\rangle}
  \nonumber \\ &=&
  \frac{i}{4}f_{P} {\int}dx_{3}\,e^{ix_{3}p_{3}{\cdot}z}
  \Big\{ {\gamma}_{5}\Big[ \!\!\not{p}_{3}\,{\phi}_{P}^{a}(x_{3})
  + {\mu}_{P}{\phi}_{P}^{p}(x_{3})
  - {\mu}_{P}(\!\not{n}_{-}\!\!\not{n}_{+}\!-\!1)\,{\phi}_{P}^{t}(x_{3})
  \Big] \Big\}_{ji}
  \label{wave-pim},
  \end{eqnarray}
  where $f_{\Upsilon}$, $f_{B_{c}}$, $f_{P}$ are
  decay constants,
  ${\mu}_{P}$ $=$ $m_{3}^{2}/(m_{u}+m_{q})$ and
  $q$ $=$ $d(s)$ for ${\pi}(K)$ meson.

  The leading twist distribution amplitudes of light
  pseudoscalar ${\pi}$, $K$ mesons are defined in
  terms of Gegenbauer polynomials \cite{jhep0605.004}:
   \begin{equation}
  {\phi}_{P}^{a}(x)=6\,x\bar{x}
   \Big\{ 1+ \sum\limits_{n=1}^{\infty}
   a_{n}^{P}\, C_{n}^{3/2}(x-\bar{x}) \Big\}
   \label{twist},
   \end{equation}
  where $\bar{x}$ $=$ $1$ $-$ $x$;
  $a_{n}^{P}$ and $C_{n}^{3/2}(z)$ are
  Gegenbauer moment and polynomials, respectively;
  $a^{\pi}_{i}$ $=$ $0$ for $i$ $=$ 1, 3, 5, ${\cdots}$
  due to the $G$-parity invariance of the pion
  distribution amplitudes.

  Because of $m_{{\Upsilon}(1S)}$ ${\simeq}$ $2m_{b}$
  and $m_{B_{c}}$ ${\simeq}$ $m_{b}$ $+$ $m_{c}$,
  both ${\Upsilon}(1S)$ and $B_{c}$ systems are nearly
  nonrelativistic, which can play the same role in
  understanding hadronic dynamics as the positronium
  and hydrogen atom in understanding the atomic
  physics \cite{ppnp61}.
  Nonrelativistic quantum chromodynamics (NRQCD)
  \cite{prd46,prd51,rmp77} and
  Schr\"{o}dinger equation can be used to describe
  their spectrum and thus one can learn about the
  interquark binding forces responsible for these
  states \cite{ann1983}.
  The eigenfunction of the time-independent
  Schr\"{o}dinger equation with scalar harmonic
  oscillator potential corresponding to the quanta
  $nL$ $=$ $1S$ is written as
   \begin{equation}
  {\phi}(\vec{k})\
  {\sim}\ e^{-\vec{k}^{2}/2{\beta}^{2}}
   \label{wave-k},
   \end{equation}
  where the parameter ${\beta}$ determines the average
  transverse momentum, i.e.,
  ${\langle}1S{\vert}\vec{k}^{2}_{\perp}{\vert}1S{\rangle}$
  $=$ ${\beta}^{2}$.
  According to the NRQCD power counting rules \cite{prd46},
  the characteristic magnitude of the momentum is order of
  $Mv$, where $M$ is the mass of the heavy quark with typical
  velocity $v$ ${\sim}$ ${\alpha}_{s}(M)$.
  Thus we will take ${\beta}$ $=$ $M{\alpha}_{s}(M)$
  in our calculation.
  Employing the substitution ansatz \cite{xiao},
   \begin{equation}
   \vec{k}^{2}\ {\to}\ \frac{1}{4} \sum\limits_{i}
   \frac{\vec{k}_{i\perp}^{2}+m_{q_{i}}^{2}}{x_{i}}
   \label{wave-kt},
   \end{equation}
  where $x_{i}$, $\vec{k}_{i\perp}$, $m_{q_{i}}$ are the
  longitudinal momentum fraction, transverse momentum,
  mass of the light valence quark, respectively,
  with the relations ${\sum}x_{i}$ $=$ $1$ and
  $\sum\vec{k}_{i\perp}$ $=$ $0$.
  Integrating out $\vec{k}_{i\perp}$ and combining with
  their asymptotic forms, one can obtain
   \begin{equation}
  {\phi}_{B_{c}}(x) = A\, x\bar{x}\,
  {\exp}\Big\{ -\frac{\bar{x}\,m_{c}^{2}+x\,m_{b}^{2}}
                     {8\,{\beta}_{2}^{2}\,x\,\bar{x}} \Big\}
   \label{wave-bc},
   \end{equation}
   \begin{equation}
  {\phi}_{\Upsilon}^{v}(x) = B\, x\bar{x}\,
  {\exp}\Big\{ -\frac{m_{b}^{2}}{8\,{\beta}_{1}^{2}\,x\,\bar{x}} \Big\}
   \label{wave-bbv},
   \end{equation}
   \begin{equation}
  {\phi}_{\Upsilon}^{t}(x) = C\, (x-\bar{x})^{2}\,
  {\exp}\Big\{ -\frac{m_{b}^{2}}{8\,{\beta}_{1}^{2}\,x\,\bar{x}} \Big\}
   \label{wave-bbt},
   \end{equation}
   where ${\beta}_{i}$ $=$ ${\xi}_{i}{\alpha}_{s}({\xi}_{i})$
   with ${\xi}_{i}$ $=$ $m_{i}/2$;
   parameters $A$, $B$, $C$ are the normalization coefficients
   satisfying the conditions
   \begin{equation}
  {\int}_{0}^{1}dx\,{\phi}_{B_{c}}(x)=1,
   \quad
  {\int}_{0}^{1}dx\,{\phi}_{\Upsilon}^{v,t}(x) =1
   \label{wave-abc}.
   \end{equation}

  \subsection{Decay amplitudes}
  \label{sec0205}
\begin{figure}[h]
  \includegraphics[width=0.99\textwidth,bb=75 620 530 720]{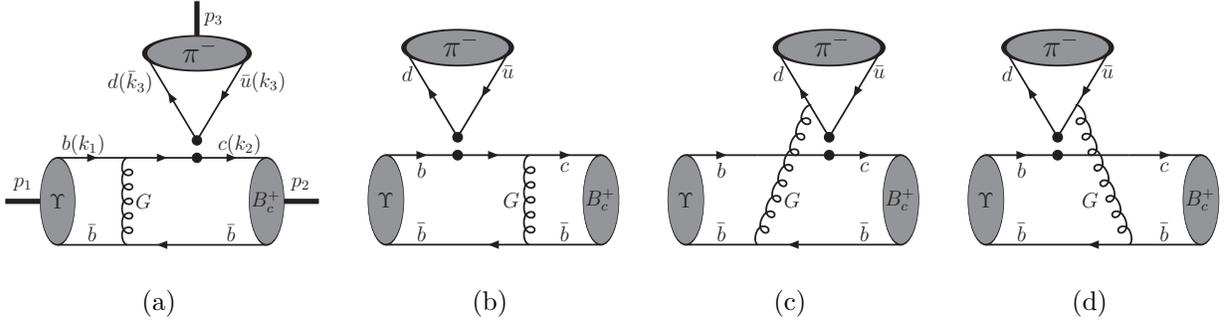}
  \caption{Feynman diagrams for the ${\Upsilon}$ ${\to}$ $B_{c}{\pi}$
   decay with the pQCD approach.}
  \label{fig1}
  \end{figure}

  The Feynman diagrams for ${\Upsilon}(1S)$ ${\to}$
  $B_{c}{\pi}$ decay are shown in Fig.\ref{fig1},
  where (a) and (b) are factorizable topology;
  (c) and (d) are nonfactorizable topology.

  With the master formula Eq.(\ref{hadronic}),
  the decay amplitudes of ${\Upsilon}(1S)$ ${\to}$ $B_{c}P$
  ($P$ $=$ ${\pi}$ and $K$) decay can be written as
   \begin{equation}
  {\cal A}({\Upsilon}(1S){\to}B_{c}P)
   = \sqrt{2}G_{F}\frac{{\pi}C_{F}}{N} V_{cb} V_{uq}^{\ast}\,
   m_{\Upsilon}^{3}\,pf_{\Upsilon}f_{B_{c}}f_{P}\!\!\!\!
   \sum\limits_{i=a,b,c,d}\!\!\!\!
  {\cal A}_{\rm Fig.\ref{fig1}(i)}
   \label{amp-all},
   \end{equation}
  where $C_{F}$ $=$ $4/3$ and the color number $N$ $=$ $3$.

  The explicit expressions of
  ${\cal A}_{\rm Fig.\ref{fig1}(i)}$
  are
   \begin{eqnarray}
   \lefteqn{ {\cal A}_{\rm Fig.\ref{fig1}(a)}\, =\,
  {\int}_{0}^{1}dx_{1} {\int}_{0}^{\infty}b_{1} db_{1}
  {\int}_{0}^{1}dx_{2} {\int}_{0}^{\infty}b_{2} db_{2}\,
  {\alpha}_{s}(t_{a})\, a_{1}(t_{a})\, E_{a}(t_{a}) }
   \nonumber \\ & & {\times}
  {\phi}_{\Upsilon}^{v}(x_{1})\,
  {\phi}_{B_{c}}(x_{2})\,
  H_{a}(x_{1},x_{2},b_{1},b_{2})\,
   \Big\{ x_{2}+r_{2}r_{b}+r_{3}^{2}\bar{x}_{2} \Big\}
   \label{amp-figa},
   \end{eqnarray}
   \begin{eqnarray}
   \lefteqn{ {\cal A}_{\rm Fig.\ref{fig1}(b)}\, =\,
  {\int}_{0}^{1}dx_{1} {\int}_{0}^{\infty}b_{1} db_{1}
  {\int}_{0}^{1}dx_{2} {\int}_{0}^{\infty}b_{2} db_{2}\,
  {\alpha}_{s}(t_{b})\, a_{1}(t_{b})\, E_{b}(t_{b}) }
   \nonumber \\ & & {\times}
  {\phi}_{B_{c}}(x_{2})\, H_{b}(x_{1},x_{2},b_{2},b_{1})\,
   \Big\{ {\phi}_{\Upsilon}^{v}(x_{1})\, \Big[
   2r_{2}r_{c}-r_{2}^{2}x_{1}-r_{3}^{2}\bar{x}_{1} \Big]
   \nonumber \\ & & +
   {\phi}_{\Upsilon}^{t}(x_{1})\,
   \Big[ 2r_{2}x_{1}-r_{c} \Big] \Big\}
   \label{amp-figb},
   \end{eqnarray}
   \begin{eqnarray}
   \lefteqn{ {\cal A}_{\rm Fig.\ref{fig1}(c)}\, =\,
  {\int}_{0}^{1}dx_{1} {\int}_{0}^{\infty} db_{1}
  {\int}_{0}^{1}dx_{2} {\int}_{0}^{\infty}b_{2} db_{2}
  {\int}_{0}^{1}dx_{3} {\int}_{0}^{\infty}b_{3} db_{3}\,
  {\delta}(b_{1}-b_{2}) }
   \nonumber \\ & & {\times}
  {\alpha}_{s}(t_{c})\, \frac{C_{2}(t_{c})}{N}\,
  E_{c}(t_{c})\, {\phi}_{B_{c}}(x_{2})\,
  {\phi}_{P}^{a}(x_{3})\,
  H_{c}(x_{1},x_{2},x_{3},b_{2},b_{3})\,
   \nonumber \\ & & {\times}
   \Big\{ {\phi}_{\Upsilon}^{v}(x_{1})
   \Big[ \frac{t\,(x_{1}-x_{3})}{m_{1}^{2}}
   +2\,r_{2}^{2}\,(x_{3}-x_{2}) \Big]
  +{\phi}_{\Upsilon}^{t}(x_{1})\,r_{2}\,(x_{2}-x_{1}) \Big\}
   \label{amp-figc},
   \end{eqnarray}
   \begin{eqnarray}
   \lefteqn{ {\cal A}_{\rm Fig.\ref{fig1}(d)}\, =\,
  {\int}_{0}^{1}dx_{1} {\int}_{0}^{\infty} db_{1}
  {\int}_{0}^{1}dx_{2} {\int}_{0}^{\infty}b_{2} db_{2}
  {\int}_{0}^{1}dx_{3} {\int}_{0}^{\infty}b_{3} db_{3}\,
  {\delta}(b_{1}-b_{2}) }
   \nonumber \\ & & {\times}
  {\alpha}_{s}(t_{d})\, \frac{C_{2}(t_{d})}{N}\,
  E_{d}(t_{d})\, {\phi}_{B_{c}}(x_{2})\,
  {\phi}_{P}^{a}(x_{3})\,
  H_{d}(x_{1},x_{2},x_{3},b_{2},b_{3})\,
   \nonumber \\ & & {\times}
   \Big\{ {\phi}_{\Upsilon}^{v}(x_{1})
   \frac{s\,(\bar{x}_{2}-x_{3})}{m_{1}^{2}}
  +{\phi}_{\Upsilon}^{t}(x_{1})\,r_{2}\,(x_{2}-x_{1}) \Big\}
   \label{amp-figd},
   \end{eqnarray}
  where ${\alpha}_{s}$ is the QCD coupling;
  $a_{1}$ $=$ $C_{1}$ $+$ $C_{2}/N$;
  $C_{1,2}$ is the Wilson coefficients;
  $r_{i}$ $=$ $m_{i}/m_{1}$.
  It can be easily seen that the nonfactorizable
  contributions ${\cal A}_{\rm Fig.\ref{fig1}(c,d)}$
  are color-suppressed with respect to the
  factorizable contributions
  ${\cal A}_{\rm Fig.\ref{fig1}(a,b)}$.

  The typical scales $t_{i}$ and the Sudakov factor $E_{i}$
  are defined as
   \begin{equation}
   t_{a(b)} = {\max}(\sqrt{-{\alpha}_{g}},
                  \sqrt{-{\beta}_{a(b)}},
                  1/b_{1},1/b_{2})
   \label{tab},
   \end{equation}
   \begin{equation}
   t_{c(d)} = {\max}(\sqrt{-{\alpha}_{g}},
                  \sqrt{{\vert}{\beta}_{c(d)}{\vert}},
                  1/b_{1},1/b_{2},1/b_{3})
   \label{tcd},
   \end{equation}
   \begin{equation}
   E_{a(b)}(t) = {\exp}\{-S_{B_{c}}(t) \}
   \label{eab},
   \end{equation}
   \begin{equation}
   E_{c(d)}(t) = {\exp}\{-S_{B_{c}}(t)-S_{P}(t) \}
   \label{ecd},
   \end{equation}
   \begin{equation}
  {\alpha}_{g} = \bar{x}_{1}^{2}m_{1}^{2}
               +  \bar{x}_{2}^{2}m_{2}^{2}
               -  \bar{x}_{1}\bar{x}_{2}t
   \label{tg},
   \end{equation}
   \begin{equation}
  {\beta}_{a} = m_{1}^{2} - m_{b}^{2}
              +  \bar{x}_{2}^{2}m_{2}^{2}
              -  \bar{x}_{2}t
   \label{tqa},
   \end{equation}
   \begin{equation}
  {\beta}_{b} = m_{2}^{2} - m_{c}^{2}
              +  \bar{x}_{1}^{2}m_{1}^{2}
              -  \bar{x}_{1}t
   \label{tqb},
   \end{equation}
   \begin{equation}
  {\beta}_{c} = x_{1}^{2}m_{1}^{2}
              +  x_{2}^{2}m_{2}^{2}
              +  x_{3}^{2}m_{3}^{2}
              -  x_{1}x_{2}t
              -  x_{1}x_{3}u
              +  x_{2}x_{3}s
   \label{tqc},
   \end{equation}
   \begin{equation}
  {\beta}_{d} = \bar{x}_{1}^{2}m_{1}^{2}
              +  \bar{x}_{2}^{2}m_{2}^{2}
              +  x_{3}^{2}m_{3}^{2}
              -  \bar{x}_{1}\bar{x}_{2}t
              -  \bar{x}_{1}x_{3}u
              +  \bar{x}_{2}x_{3}s
   \label{tqd},
   \end{equation}
   \begin{equation}
   S_{B_{c}}(t) = s(x_{2},p_{2}^{+},1/b_{2})
                +2{\int}_{1/b_{2}}^{t}\frac{d{\mu}}{\mu} {\gamma}_{q}
   \label{sbc},
   \end{equation}
   \begin{equation}
   S_{P}(t) = s(x_{3},p_{3}^{+},1/b_{3})
            + s(\bar{x}_{3},p_{3}^{+},1/b_{3})
            +2{\int}_{1/b_{3}}^{t}\frac{d{\mu}}{\mu}{\gamma}_{q}
   \label{sp},
   \end{equation}
  where ${\alpha}_{g}$ and ${\beta}_{i}$ are the virtuality
  of the internal gluon and quark, respectively;
  ${\gamma}_{q}$ $=$ $-{\alpha}_{s}/{\pi}$ is the quark
  anomalous dimension; the expression of $s(x,Q,1/b)$ can
  be found in the appendix of Ref.\cite{pqcd1}.

  The scattering functions $H_{i}$ in the subamplitudes
  ${\cal A}_{\rm Fig.\ref{fig1}(i)}$
  are defined as
   \begin{equation}
   H_{a(b)}(x_{1},x_{2},b_{i},b_{j})\, =\,
   K_{0}(\sqrt{-{\alpha}_{g}}b_{i})
   \Big\{ {\theta}(b_{i}-b_{j})
   K_{0}(\sqrt{-{\beta}}b_{i})
   I_{0}(\sqrt{-{\beta}}b_{j})
   + (b_{i}{\leftrightarrow}b_{j}) \Big\}
   \label{hab},
   \end{equation}
   \begin{eqnarray}
   H_{c(d)}(x_{1},x_{2},x_{3},b_{2},b_{3}) &=&
   \Big\{ {\theta}(-{\beta}) K_{0}(\sqrt{-{\beta}}b_{3})
  +\frac{{\pi}}{2} {\theta}({\beta}) \Big[
   iJ_{0}(\sqrt{{\beta}}b_{3})
   -Y_{0}(\sqrt{{\beta}}b_{3}) \Big] \Big\}
   \nonumber \\ &{\times}&
   \Big\{ {\theta}(b_{2}-b_{3})
   K_{0}(\sqrt{-{\alpha}_{g}}b_{2})
   I_{0}(\sqrt{-{\alpha}_{g}}b_{3})
   + (b_{2}{\leftrightarrow}b_{3}) \Big\}
   \label{hcd},
   \end{eqnarray}
  where $J_{0}$ and $Y_{0}$ ($I_{0}$ and $K_{0}$) are the
  (modified) Bessel function of the first and second kind,
  respectively.

  \section{Numerical results and discussion}
  \label{sec03}

  In the rest frame of the ${\Upsilon}(1S)$ particle,
  branching ratio for the ${\Upsilon}(1S)$ ${\to}$
  $B_{c}P$ weak decays can be written as
   \begin{equation}
  {\cal B}r({\Upsilon}(1S){\to}B_{c}P)\ =\ \frac{1}{12{\pi}}\,
   \frac{p}{m_{{\Upsilon}}^{2}{\Gamma}_{{\Upsilon}}}\,
  {\vert}{\cal A}({\Upsilon}(1S){\to}B_{c}P){\vert}^{2}
   \label{br},
   \end{equation}
  where the decay width ${\Gamma}_{\Upsilon}$ $=$
  $54.02{\pm}1.25$ keV \cite{pdg}.

  The values of input parameters are listed as follows.

  (1) Wolfenstein parameters \cite{pdg}:
    $A$ $=$ $0.814^{+0.023}_{-0.024}$ and
    ${\lambda}$ $=$ $0.22537{\pm}0.00061$.

  (2) Masses of mesons \cite{pdg}:
    $m_{B_{c}}$ $=$ $6275.6{\pm}1.1$ MeV and
    $m_{{\Upsilon}(1S)}$ $=$ $9460.30{\pm}0.26$ MeV.

  (3) Masses of quarks \cite{pdg}:
    $m_{c}$ $=$ $1.67{\pm}0.07$ GeV and
    $m_{b}$ $=$ $4.78{\pm}0.06$ GeV.

  (4) Gegenbauer moments at the scale of ${\mu}$ $=$ 1 GeV:
    $a_{2}^{\pi}$ $=$ $0.17{\pm}0.08$ and
    $a_{4}^{\pi}$ $=$ $0.06{\pm}0.10$ \cite{prd83}
    for twist-2 pion distribution amplitudes,
    $a_{1}^{K}$ $=$ $0.06{\pm}0.03$ and
    $a_{2}^{K}$ $=$ $0.25{\pm}0.15$ \cite{jhep0605.004}
    for twist-2 kaon distribution amplitudes.

  (5) Decay constants:
    $f_{\pi}$ $=$ $130.41{\pm}0.20$ MeV \cite{pdg},
    $f_{K}$ $=$ $156.2{\pm}0.7$ MeV \cite{pdg},
    $f_{B_{c}}$ $=$ $489{\pm}5$ MeV \cite{fbc}.
    As for the decay constant $f_{{\Upsilon}}$,
    one can use the definition of decay constant
    $f_{V}$ for vector meson $V$ with mass $m_{V}$
    and polarization vector ${\epsilon}_{V}$,
   \begin{equation}
  {\langle}0{\vert}\bar{\psi}{\gamma}^{\mu}{\psi}{\vert}V{\rangle}
   = f_{V}m_{V}{\epsilon}_{V}^{\mu}
   \label{fvv1}.
   \end{equation}
   The decay constant $f_{V}$ is related to the experimentally
   measurable leptonic branching ratio:
   \begin{equation}
  {\Gamma}(V{\to}{\ell}^{+}{\ell}^{-})\, =\,
   \frac{4{\pi}}{3}{\alpha}_{\rm QED}^{2}Q_{q}^{2}
   \frac{f_{V}^{2}}{m_{V}}
   \sqrt{ 1-2\frac{m_{\ell}^{2}}{m_{V}^{2}} }
   \Big\{ 1+2\frac{m_{\ell}^{2}}{m_{V}^{2}} \Big\}
   \label{fvv2},
   \end{equation}
  where ${\alpha}_{\rm QED}$ is the fine-structure
  constant, $m_{\ell}$ is the lepton mass, $Q_{q}$ is the
  electric charge of the quark in the unit of ${\vert}e{\vert}$,
  and $Q_{b}$ $=$ $-1/3$ for the bottom quark.
  The experimental measurements on leptonic
  ${\Upsilon}(1S)$ decays give the weighted average
  decay constant $f_{{\Upsilon}}$ $=$
  $(676.4{\pm}10.7)$ MeV (see Table.\ref{tab:fbb}).

   \begin{table}[h]
   \caption{Branching ratios for leptonic ${\Upsilon}(1S)$
   decays and decay constants $f_{{\Upsilon}}$, where
   the last column is the weighted average,
   and errors come from mass, width and
   branching ratios.}
   \label{tab:fbb}
   \begin{ruledtabular}
   \begin{tabular}{lccc}
   decay mode & branching ratio &
   \multicolumn{2}{c}{decay constant} \\ \hline
    ${\Upsilon}(1S)$ ${\to}$ $e^{+}e^{-}$
  & $(2.38{\pm}0.11)\%$
  & $(664.2{\pm}23.1)$ MeV & \\
    ${\Upsilon}(1S)$ ${\to}$ ${\mu}^{+}{\mu}^{-}$
  & $(2.48{\pm}0.05)\%$
  & $(677.9{\pm}14.7)$ MeV
  & $(676.4{\pm}10.7)$ MeV \\
    ${\Upsilon}(1S)$ ${\to}$ ${\tau}^{+}{\tau}^{-}$
  & $(2.60{\pm}0.10)\%$
  & $(683.3{\pm} 21.1)$ MeV &
  \end{tabular}
  \end{ruledtabular}
  \end{table}
   \begin{table}[h]
   \caption{Branching ratios for the ${\Upsilon}(1S)$ ${\to}$
   $B_{c}{\pi}$, $B_{c}K$ decays, where the results of Refs. \cite{ijma14,adv2013,691261}
   are calculated with the coefficient $a_{1}$ $=$ $1.05$.
   The uncertainties of the last column come from the CKM
   parameters, the renormalization scale ${\mu}$ $=$ $(1{\pm}0.1)t_{i}$,
   masses of $b$ and $c$ quarks, hadronic parameters (decay
   constants and Gegenbauer moments), respectively.}
   \label{tabbr}
   \begin{ruledtabular}
  \begin{tabular}{l|c|c|c|c}
    & Ref. \cite{ijma14} & Ref. \cite{adv2013}
    & Ref. \cite{691261} & this work \\ \hline
    $10^{11}{\times}{\cal B}r({\Upsilon}(1S){\to}B_{c}{\pi})$
  & $6.91$ & $2.8$ & $5.03$
  & $7.04^{+0.48+0.80+0.83+0.47}_{-0.46-0.52-0.98-0.44} $\\ \hline
    $10^{12}{\times}{\cal B}r({\Upsilon}(1S){\to}B_{c}K)$
  & $5.03$ & $2.3$ & $3.73$
  & $5.41^{+0.40+0.63+0.64+0.39}_{-0.38-0.41-0.74-0.37} $
  \end{tabular}
  \end{ruledtabular}
  \end{table}

  If not specified explicitly, we will take their
  central values as the default inputs.
  Our numerical results on the $CP$-averaged branching ratios
  for the ${\Upsilon}(1S)$ ${\to}$ $B_{c}{\pi}$, $B_{c}K$ decays
  are displayed in Table \ref{tabbr},
  where the uncertainties come from the CKM parameters,
  the renormalization scale ${\mu}$ $=$ $(1{\pm}0.1)t_{i}$,
  masses of $b$ and $c$ quarks, hadronic parameters
  (decay constants and Gegenbauer moments), respectively.
  The following are some comments.

  (1)
  The pQCD's results on branching ratios for the
  ${\Upsilon}(1S)$ ${\to}$ $B_{c}{\pi}$, $B_{c}K$ decays
  have the same magnitude as those of
  Refs. \cite{ijma14,adv2013,691261}.
  The estimation of Refs. \cite{ijma14,adv2013} is based
  on the NF approximation,
  where nonfactorizable corrections to HME are not considered,
  and the form factors for the transition between ${\Upsilon}(1S)$
  and $B_{c}$ mesons are calculated with the heavy quark
  effective theory in Ref. \cite{ijma14} and the Wirbel-Stech-Bauer
  \cite{bsw} model in Ref. \cite{adv2013}, respectively.
  The coefficient $a_{1}$ containing the NLO nonfactorizable
  contributions to HME are used in Ref. \cite{691261} within the
  QCDF framework, where the
  form factors are written as the overlap integrals
  of nonrelativistic wave functions for ${\Upsilon}(1S)$
  and $B_{c}$ mesons based on the Wirbel-Stech-Bauer model.
  Compared with the NF and QCDF approach,
  there are more contributions from the nonfactorizable
  decay amplitudes ${\cal A}_{\rm Fig.\ref{fig1}(c,d)}$
  with the pQCD approach.
  This may be why the pQCD's results are slightly
  larger than previous ones.

  (2)
  Because the CKM factors
  ${\vert}V_{cb}V_{us}^{\ast}{\vert}$ $<$
  ${\vert}V_{cb}V_{ud}^{\ast}{\vert}$,
  there is a relation between branching ratios,
  ${\cal B}r({\Upsilon}(1S){\to}B_{c}{\pi})$ $>$
  ${\cal B}r({\Upsilon}(1S){\to}B_{c}K)$.

  (3)
  Branching ratio for the ${\Upsilon}(1S)$ ${\to}$ $B_{c}{\pi}$
  decay can reach up to $10^{-11}$. So
  the ${\Upsilon}(1S)$ ${\to}$ $B_{c}{\pi}$ decay should be
  sought for with high priority and first observed at the
  running LHC and forthcoming SuperKEKB.
  For example, the ${\Upsilon}(1S)$ production cross
  section in p-Pb collision can reach up to a few ${\mu}b$
  with the LHCb \cite{jhep1407} and ALICE \cite{plb740}
  detectors at LHC.
  Over $10^{11}$ ${\Upsilon}(1S)$ particles per 100 $fb^{-1}$
  data collected at LHCb and ALICE  are in principle available,
  corresponding to a few tens of
  ${\Upsilon}(1S)$ ${\to}$ $B_{c}{\pi}$ events.

  (4)
  There are many uncertainties on our results.
  Other factors, such as the contributions of higher
  order corrections to HME, relativistic effects
  and so on, which are not considered here,
  deserve the dedicated study.
  Our results just provide an order of magnitude estimation.

  \section{Summary}
  \label{sec04}
  The ${\Upsilon}(1S)$ weak decay is legal within
  the standard model, although branching ratio is tiny
  compared with the strong and electromagnetic decays.
  With the potential prospects of the ${\Upsilon}(1S)$
  at high-luminosity dedicated heavy-flavor factories,
  the bottom-changing ${\Upsilon}(1S)$ ${\to}$ $B_{c}{\pi}$,
  $B_{c}K$ weak decays are studied with the pQCD approach.
  It is found that with the nonrelativistic wave functions
  for ${\Upsilon}(1S)$ and $B_{c}$ mesons, branching
  ratios for the ${\Upsilon}(1S)$ ${\to}$ $B_{c}{\pi}$,
  $B_{c}K$ decays have the same order as previous works,
  and ${\cal B}r({\Upsilon}(1S){\to}B_{c}{\pi})$
  ${\gtrsim}$ $10^{-11}$.
  The color-favored and CKM-favored ${\Upsilon}(1S)$
  ${\to}$ $B_{c}{\pi}$ decay might be detectable in
  future experiments.

  \section*{Acknowledgments}
  We thank Professor Dongsheng Du (IHEP@CAS) and Professor
  Yadong Yang (CCNU) for helpful discussion.
  We thank the referees for their constructive comments.
  The work is supported by the National Natural Science Foundation
  of China (Grant Nos. 11475055, 11275057, U1232101 and 11347030)
  and the Program of Education Department of Henan Province
  (Grant No. 14HASTIT037).

  
  \end{document}